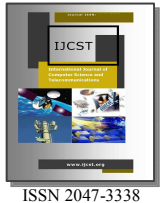

# Priority Based Pre-emptive Task Scheduling for Android Operating System

Deepali Kayande[1] and Urmila Shrawankar[2]

[1,2]Department of Computer Science and Engineering, G. H. Raisoni College of Engineering, Nagpur-India
kayande.deepali@ieee.org; urmila@ieee.org

*Abstract*— Android mobile operating system which is based on Linux Kernel 2.6, has open source license and adaptability to user driven applications. As all other operating systems it has all the basic features like process scheduling, memory management, process management etc associated with it. Any mobile platform works smoothly when the process scheduling is performed in a proper way. Ideal platform is that in which no resource conflict occurs. Thus scheduling in every manner is essential for the operating system to adapt itself with the requirement of a particular application. In this paper, priority based pre-emptive task scheduling is proposed for the SMS application. The idea is to define High priority to required contacts, for ex. Contact numbers of parents or teachers will be given High priority. If in case, any SMS from these High priority contacts is received, the application would flash the SMS on the active screen and redirect this High priority SMS to the Priority Inbox.

*Index Terms*— Android OS, Mobile Operating System, Pre-emptive Task Scheduling and SMS Application

## I. INTRODUCTION

OPERATING System is considered as efficient when its throughput rate is high. Task scheduling, memory management are the few aspects which makes it possible to increase the response time of operating systems. Mobile operating systems are the embedded devices which are designed for the specific use and are expected to meet the specific time deadlines for completing the tasks. For this, the response time of the important tasks is most important which can be achieved using the priority based task scheduling. This paper puts forth the idea of applying the pre-emptive based task scheduling for SMS application. SMS is the technology that enables the sending and receiving of messages over the network via exchange of text files.

Now-a-days, SMS files are extended to carry binary data viz. ringtones, pictures, business cards, etc. Thus developing a technique which will separate important SMS files into a different priority inbox from the default one helps in better search and better storage space utilization. This idea has been implemented taking into account the working of Gmail's Priority Inbox. In this setting a user can separate out the mails present in the mail's in-box in different categories viz. starred, unread, important and read. Thus this separating of mails helps in better categorization of the bulky mails present in the in-box. In this project the idea is to define High priority to required contacts from the contact list. Defining High priority will enable the application to flash the important SMS on the active screen thereby alerting the user.

The paper is divided into eight sections. Section II of the paper deals with Android architecture. Section III explains the priority based pre-emptive task scheduling. Section IV deals with the experimental steps for the proposed technique, algorithm and flowchart. The snippets for developing the Priority Manager Application is discussed in Section V which is followed with the experimental results described in Section VI. Section VII tells about the advantages of the proposed technique over the existing system and finally the observations and conclusion are proposed in Section VIII.

## II. ANDROID ARCHITECTURE

Following are the features present in Android architecture [9]:

i) *Application framework:* this enables the reuse and replacement of components
ii) *Dalvik virtual machine:* it is optimized for mobile devices
iii) *Integrated browser:* it is based on the open source Web Kit engine

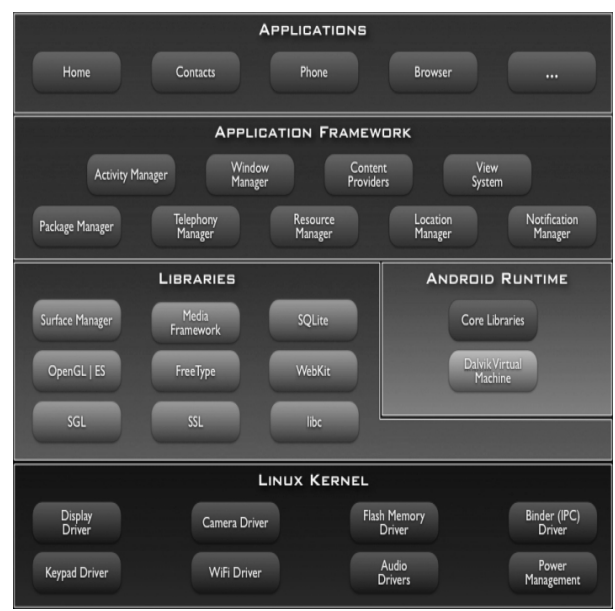

Fig 1: Android Architecture [9]

iv) *Optimized graphics:* it is powered by a custom 2D graphics library; 3D graphics based on the OpenGL ES 1.0 specification (hardware acceleration optional)
v) *SQLite:* it is used the for structured data storage
vi) *Media support:* it is used for common audio, video, and still image formats (MPEG4, H.264, MP3, AAC, AMR, JPG, PNG, GIF)
vii) *GSM Telephony*
viii) *Bluetooth, EDGE, 3G and WiFi*
ix) *Camera, GPS, compass and accelerometer*
x) *Rich development environment:* it includes a device emulator, tools for debugging, memory and performance profiling, and a plugin for the Eclipse IDE.

### III. PRE-EMPTIVE TASK SCHEDULING

The key concept present in any operating system which allows the system to support multitasking, multiprocessing, etc. is Task Scheduling [1]. Task Scheduling is the core which refers to the way the different processes are allowed to share the common CPU. Scheduler and dispatcher are the softwares which help to carry out this assignment [2]. Android operating system uses O (1) scheduling algorithm as it is based on Linux Kernel 2.6. Therefore the scheduler is names as Completely Fair Scheduler as the processes can schedule within a constant amount of time, regardless of how many processes are running on the operating system [6], [7].

Pre-emptive task scheduling involves interrupting the low priority tasks when high priority tasks are present in the queue. This scheduling is particularly used for mobile operating system as the CPU utilization is medium, turnaround time and response time is high. Mobile phones are required to meet specific time deadlines for the tasks to occur.

### IV. EXPERIMENTAL STEPS

In this section the steps implemented to achieve the fixed priority pre-emptive task scheduling is described. Activities, objects and Classes provided by the Developer's site for Android are used for the implementation purpose. Activity like ListActivity, object like ListView and class like BroadcastReceiver have been used [9], [10].

The steps for the proposed technique are as follows:

i) The application would first register to a class called ListActivity that displays a list of items by binding to a data source such as an array or cursor, and exposes event handlers when the user selects an item. ListActivity hosts a ListView object that can be bound to different data sources, typically either an array or a Cursor holding query results.
ii) Now, the application would show an interface by which the user can set priority to certain contacts as 'High', and others will be kept as default.
iii) In case, there is a SMS from a high priority person while some activity is running on the phone, that message being sent by the High prioritized person, will get flashed on the active screen.
iv) At the same time this particular message will be redirected and stored in the PriorityManager inbox as well as in the default inbox.
v) If SMS is received from the contacts which are the default priority contacts, then no running activity will be disturbed and the SMS will go to the default inbox.
vi) Android's BroadcastReceiver Class would be used for this purpose.
vii) The BroadcastReceiver class Provides access to information about the SMS services on the device.
viii) Applications can use the methods in this class to determine SMS services.
ix) A file will be created which will act as the intermediate messenger between the different classes as well as different functions.

*A. Algorithm*

1. Start the SMS application
2. Register incoming SMS
3. SMS received by BroadcastReceiver
4. Check the contact number for priority

   if High priority, then

   Flash SMS on the active screen and take backup in PriorityManager inbox

   else

   redirect to inbox.

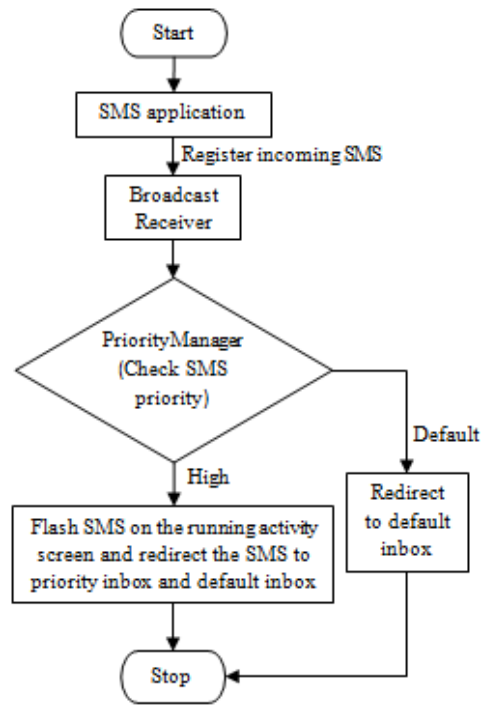

Fig. 2: Flowchart for the proposed technique

### V. CODE DESCRIPTION

In order to achieve the pre-emptive task scheduling for the SMS application, various JAVA files are created which consists of different methods and classes [9], [10].

Methods like SmsReceiver(), getMessageId(), ViewMessages(), onCreate(), onCreateOptionsMenu(), etc are used.

Following is the snippet for onReceive method which takes the required action when any SMS file is received.

*public void* onReceive(Context context,  Intent intent){

   *for (int* i = 0; i < msgs.length; i++) {

   msgs[i] = SmsMessage.createFromPdu((*byte*[]) pdus[i]);

   str += msgs[i].getOriginatingAddress();

   str += " :";

   str += msgs[i].getMessageBody().toString();

   }
}

Following code gives idea about the ViewMessages method which is used to view the SMS received from the High Priority contacts and which are stored in the Priority Inbox.

*public class* ViewMessages *extends* ListActivity{

 *while* (cur.moveToNext()){
*if*(cur.getString(2).contains(CentralRepository.CURRENT_ CONTACT)){

         lstMessages.add(cur.getString(11));

           }
     }
}

Following is the snippet that explains the action performed when the SMS to be viewed from the Priority Inbox is selected. When the SMS is selected the text from the SMS file is flashed on the screen.

lv.setOnItemClickListener(*new* OnItemClickListener() {

   *public void* onItemClick(AdapterView<?> arg0, View arg1, *int* arg2,    *long* arg3) {

      TextView txt = (TextView) arg1;
Toast.makeText(app_context,txt.getText(),Toast.LENGTH_ SHORT).show();

       }
    }
}

## VI.  EXPERIMENTAL RESULTS

From the available list of contacts in mobile phones, few contacts which are important can be set as High priority contacts. Android's Emulator has been used for the experimental purpose. Emulator is a virtual mobile device that runs on the computer. The emulator allows developing and testing Android applications without using a physical device. In the following figures, the numbers like 5554, 5556 etc. resemble the 10 digit contact numbers present in the mobile phone's contact list.

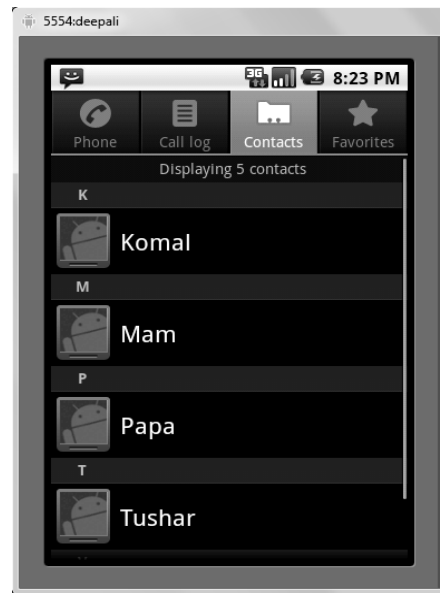

Fig. 3: Contact List saved in mobile phone

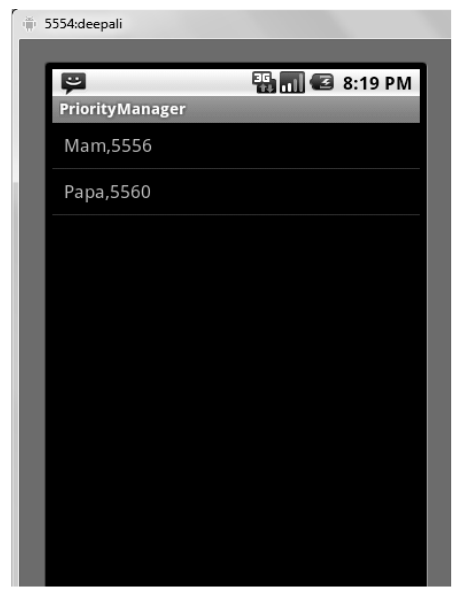

Fig. 4: PriorityManager showing the list of High Priority Contacts which are selected from the list of contacts present in the phone memory

When any SMS file is received from the Default priority contact then no interruption occurs on the working task. Also this received SMS file is stored in the default inbox present in mobile phone.

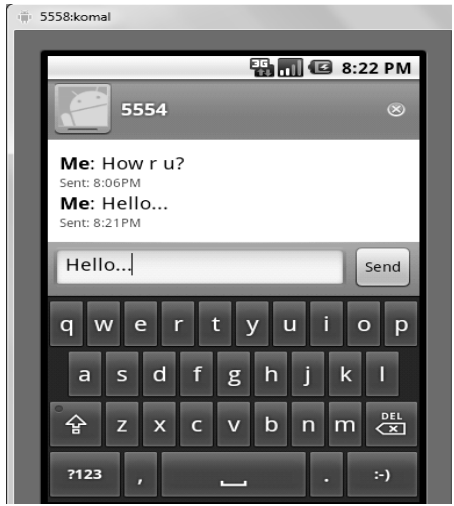

Fig. 5: Default Priority contact typing the SMS which is to be sent

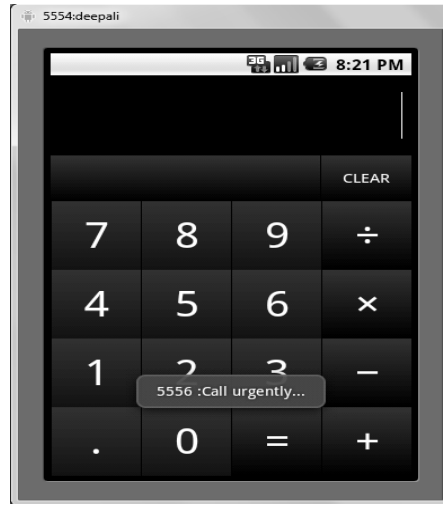

Fig. 8: The ongoing task is interrupted when the SMS is received from High Priority Contact. This SMS is flashed on the screen so as to notify the user. This High priority SMS is stored in default inbox as well as in the priority inbox designed for storing the important SMS received from the High priority contacts.

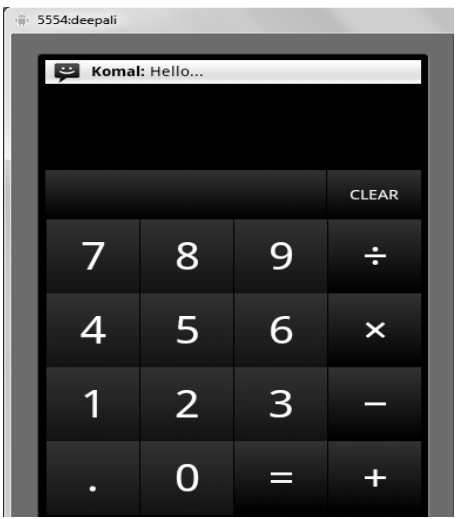

Fig. 6: No interruption on the ongoing task when the SMS file sent by the Default priority contact is received. This SMS file is redirected to the default inbox present in the phone.

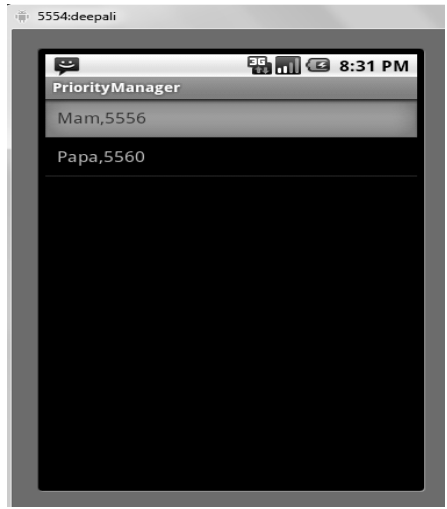

Fig. 9: Selecting the High priority contact number (Mam) to read the SMS which is stored in the priority inbox

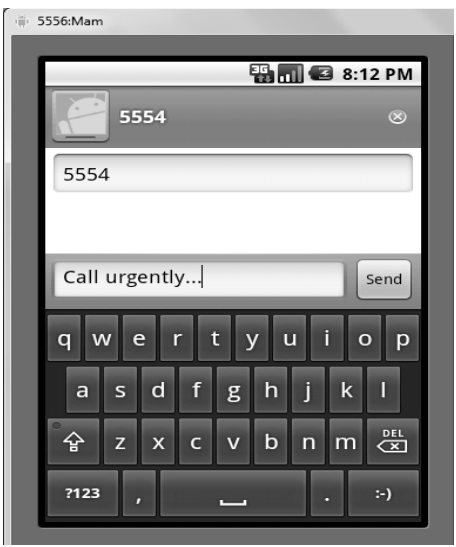

Fig. 7: High Priority contact typing the SMS which is to be sent

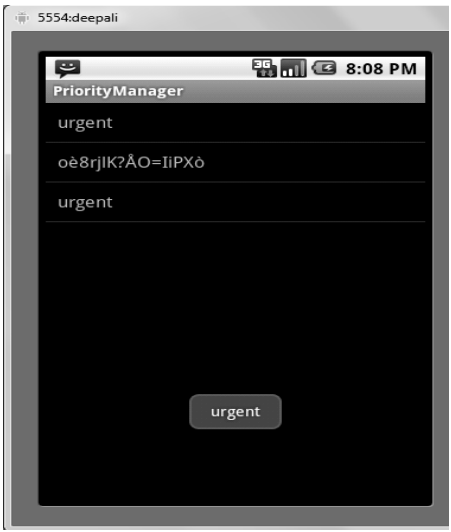

Fig. 10: Selected SMS from the priority inbox flashing on the screen

## VII. ADVANTAGES OVER EXISTING SYSTEM

Various scheduling techniques are available for the operating systems. O (1) scheduling algorithm is used for the Linux Kernel [6], [7].

But Pre-emptive task scheduling is not available for SMS application in any mobile operating system. This technique is innovative and useful. It generally happens that important SMS at times get merged in the inbox and it takes lot of time to search the important SMS containing the important data. Thus designing the high priority inbox to segregate such important SMS solves the purpose.

The proposed technique works as an add-on for Android Mobile Operating System. This technique facilitates to categorize the contacts present in the mobile phone's contact lists as High or Default. Thus using this technique helps in reducing the search time for important SMS files received from the High priority contacts. Categorization of important SMS is achieved by redirecting them to the Priority Inbox. Thus this technique is useful where the amount of SMS files to be dealt with is plenty.

## VIII. OBSERVATIONS AND CONCLUSION

Proposed application aims towards better utilization of the contacts list present in the mobile phones. It is rarely thought of categorizing the important contacts as 'High' priority contacts and keeping other contacts are 'Default'. This idea has been thought of and has been implemented in this application. A very basic and day to day used SMS application has been targeted. This is done because SMS is most widely used after the call application.

The proposed technique is used for the SMS files for the Android Mobile Operating System. This feature of High Priority Contacts and the pre-emptive task scheduling can be extended for the Call application present on mobile phones. Also mobile platforms like Symbian, Bada, Windows, etc can be tested.